\documentclass[usenatbib,fleqn,twocolumn]{mnras}
\usepackage{float}
\usepackage{mathptmx}
\usepackage{amsmath,amssymb,amsfonts,amsbsy} 
\usepackage[pdftex]{graphicx} 
\usepackage[usenames,dvipsnames]{color}
\usepackage{natbib}
\usepackage{ulem}
\DeclareSymbolFont{cmletters}{OML}{cmm}{m}{it}
\DeclareMathSymbol{v}{\mathalpha}{cmletters}{"76}

\definecolor{MyDarkBlue}{rgb}{0,0.1,0.7}
\hypersetup{pdfborder={0 0 0},colorlinks,breaklinks=true,
  urlcolor={blue},citecolor={MyDarkBlue},linkcolor={magenta}}

\newcommand{\ie}{i.e.,~}
\newcommand{\eg}{e.g.,~}

\title[Evolution of PNSs to pulsars, magnetars and CCOs]{Evolution of proto-neutron stars to pulsars, magnetars and central compact objects}

\author[Bak{\i}r \& Ek{\c s}i]{{\.I}rem Bak{\i r}$^1$ and Kaz{\i}m Yavuz Ek\c{s}i$^1$ \\  
 $^1$Istanbul Technical University,
  Faculty  of Science  and  Letters,  Physics Engineering  Department,
  34469,  Istanbul, Turkey, \href{mailto:bakirir@itu.edu.tr}{bakirir@itu.edu.tr}, \href{mailto:eksi@itu.edu.tr}{eksi@itu.edu.tr}}

\begin{document}

\maketitle

\begin{abstract}
Some young neutron stars, the magnetars, have ultra-strong magnetic fields, yet their inferred birth rate is comparable to the core-collapse supernova rate, challenging scenarios that require rare, extreme conditions. We propose that this discrepancy can be reconciled if both pulsars and magnetars pass through a dynamo process during the proto-neutron star (PNS) phase. We employ a shear-driven $\alpha$--$\Omega$ dynamo model that includes PNS contraction. The dynamo generically produces toroidal-dominated fields set mainly by the $\Omega$-effect. The evolution of the poloidal field is first dominated by flux conservation during collapse and then by the $\alpha$-effect. The saturated toroidal field depends strongly on the initial value of the shear, with a threshold at $q_0 \simeq 0.23$; below this, the poloidal field remains near the value obtained by the flux-conservation ($\approx 2.5\times10^{10}\,{\rm G}$). For the shortest initial periods, the model leads to magnetar-like strengths ($B_{\rm p} \simeq 10^{15}\,{\rm G}$, $B_\phi \simeq 10^{16}\,{\rm G}$), while for the slower rotators it yields ordinary pulsar fields ($B_{\rm p} \simeq 10^{12}\,{\rm G}$, $B_\phi \simeq 10^{14}\,{\rm G}$). We also argue that the central compact objects can acquire toroidal fields amplified solely by the $\Omega$-effect; lacking the $\alpha$-effect, their poloidal fields are not shaped by the dynamo effect.
\end{abstract}


\section{Introduction}
\label{sec:intro}

Discussions about the origin of the strong magnetic fields of neutron stars have a long history \citep{spr09}, rekindled by the identification of magnetars, \ie neutron stars with super-strong magnetic fields \citep{kas17}. According to the ``fossil field'' hypothesis \citep{wol64,gin64,rud72}, the conservation of magnetic flux during core-collapse would lead to the amplification of magnetic fields. This is commonly assumed to be the origin of the magnetic fields of conventional neutron stars. One line of argument also invokes the ``fossil field'' hypothesis for the origin of magnetic fields of magnetars \citep{fer06,fer08}, suggesting that they descend from the strong-field tail of the progenitor distribution. \citet{mak+21} tested this hypothesis using population synthesis methods and found that the model can not explain the distribution of the inferred magnetic fields of the neutron star population because the number of high-field progenitors is not sufficiently large.

Another line of argument for the origin of neutron star magnetic fields invokes the dynamo action \citep{rud73}, and this is widely accepted as the origin of magnetic fields in magnetars \citep{dun92,tho93}. Numerical simulations \citep{bon+03,bon+05,bon+06,nas+08,rhe05,gep06,ray+20,lan+21} suggest that the dynamo action could indeed produce the magnetar fields if the neutron star begins its life with a period of a few milliseconds \citep[see][for a review]{bon08}.
The dynamo process is active at the proto-neutron star \citep[PNS;][]{pra+97,pon+99,cam+16} stage with initial radius $R_0 \sim 40\,{\rm km}$ lasting about $t \sim 30-40\,{\rm s}$ \citep{ray+20}.

The rather extreme source parameters invoked by the magnetar formation scenarios (very massive and magnetized progenitors in the case of the fossil fields and very rapidly rotating neutron stars in the case of the dynamo models) seem to be at odds with the relatively high birth rates of magnetars. The core-collapse rate in the Galaxy is $1.9\pm 1.1 \, {\rm century^{-1}}$, as implied by the high-spectral-resolution measurements of $^{26}$Al emission at $1808.65\,{\rm keV}$ \citep{die+06}. A more recent estimate is $1.63 \pm 0.46\,{\rm century^{-1}}$ \citep{roz+21}. The birth rate of magnetars is estimated to be $0.3\, {\rm century^{-1}}$ \citep{kea08} and $0.23-2\,{\rm century^{-1}}$ \citep{ben+19}. This is only an order of magnitude smaller or comparable to the core collapse rate, suggesting that magnetar formation is not as rare as the formation scenarios suggest. 

Further evidence for the ordinary magnetar formation parameters comes from X-ray observations of supernova remnants, which provide no clue that the formation of magnetars involved a greater energy input than the formation of ordinary neutron stars \citep{vin06}. Moreover, magnetars appear to have typical space velocities \citep{del+12,ten+13} rather than being on the high end tail of the space velocity distribution, as some magnetar models claim \citep{dun92,tho93}. Any model for magnetar formation should reconcile the extreme parameter requirements with the high formation rates, an issue highlighted by \citet{mer+15}.

Neutron stars descend from O- and B-type stars which are massive ($M_* \sim 10 M_{\sun}$) 
and large $R_* \sim 10^{11}~{\rm cm}$. Such stars typically have surface magnetic fields of order $\sim 1\,{\rm G}$, and only a small fraction of them has fields reaching $B_{\max} \sim 10^4\,{\rm G}$ \citep{don09}. 
Textbook arguments employ magnetic flux conservation as if the whole star collapses to 
the radius of the neutron star, $B_{\rm NS} = B_* (R_*/R_{\rm NS})^2$, corresponding to a factor $10^{10}$ amplification of the field which seems to provide even the magnetar-like 
fields $B_{\rm NS}\sim 10^{14}\,{\rm G}$. The iron core, the collapsing component of the star immediately before the collapse, has a radius about $R_{\rm core} \simeq 3\,000\,{\rm km}$ \citep{suk+16}. The magnetic field inferred from the seismic observations suggests that the field near the core is about $B_{\rm core} = 5\times 10^5\,{\rm G}$ \citep{lec+22}. Thus, as recently argued by \citet{lan+21}, this would result in $B_{\rm NS} = B_{\rm core} (R_{\rm core}/R_{\rm NS})^2$, which corresponds to a factor of $10^{5}$ amplification, leading to $B_{\rm NS} \sim {\rm a~few}\times 10^{10}\,{\rm G}$. Obviously, this field is $1-2$ orders of magnitude lower than typical pulsar fields ($B_{\rm p} \sim 10^{12}~{\rm G}$), implying a necessity for dynamo action for typical pulsars, though not as vigorous as that proceeding in magnetars. Interestingly, such fields of $B_{\rm NS} \sim 10^{10}\,{\rm G}$ are of the order of the magnetic fields of central compact objects \citep[CCOs;][]{got07,hal+07,got09,del17}. These arguments suggest that the fossil field arguments can only address the dipole magnetic fields of CCOs. Note also that these objects could also have stronger small-scale magnetic fields \citep{gou+20,igo+21CCO}. 

Since the core does not directly collapse into a neutron star but passes through the stage of a PNS with an initial radius of $R_0 \sim  40\,{\rm km}$, the flux conservation implies that the PNS inherits a magnetic field about
\begin{equation}
B_{\rm PNS} = B_{\rm core} \left( \frac{R_{\rm core}}{R_0}\right)^2 \simeq 3\times 10^9\,{\rm G}
\label{eq:flux_cons}
\end{equation}
which then will be enhanced, upon collapse to $R_{\infty} =12~{\rm km}$, by $R_0/R_{\infty}=(40~{\rm km}/12~{\rm km})^2 \simeq 11$ times  reaching $B_{\rm p} \sim {\rm a~few} \times 10^{10}~{\rm G}$. We note that, this is an over-estimate, and in the presence of turbulent reconnection driven flux loss, a lower value will be reached.

The results of \citet{ray+20} suggest that the dynamo-generated magnetic fields depend on the initial periods, slower initial periods resulting in less vigorous dynamo action and weaker magnetic fields. By analyzing the results of three-dimensional core-collapse simulations, \citet{whi+22} argued that the pulsar/magnetar dichotomy arises from the different rotation rates of the cores of their progenitors. \citet{bar+22} brought into attention the role of fallback matter in spinning up the PNS to induce vigorous dynamo action to lead to magnetar field strengths. 

Given the existence of low-magnetic field ``magnetars'' \citep{rea+10,rea+13,rea+14} with dipole magnetic fields as low as a few $10^{12}$~G, the parameter that makes magnetars different from typical pulsars is most likely their relatively stronger toroidal magnetic fields which generically is attributed to the dynamo process the magnetars pass through at the PNS stage.
In this work, we assume that both ordinary neutron stars and magnetars have dynamo-generated fields and study the parameter space to find what leads to the differences in the poloidal and toroidal magnetic fields. We use and suitably extend the one-dimensional dynamo model \citep{wic+14} originally proposed for magnetic field generation in white dwarf mergers. We seek what initial parameters will lead to the different toroidal field strengths at the end of the dynamo stage to determine whether the PNS becomes a magnetar ($B_{\phi} \gtrsim 10^{15}\,{\rm G}$) or a typical neutron star ($B_{\phi} \sim 10^{14}\,{\rm G}$).

The organization of the paper is as follows. In \S~\ref{sec:model}, we introduce the shear-driven dynamo model of \citet{wic+14}, adopt it to the PNS parameters, and add the effects of changing radius and gravitational mass. In \S~\ref{sec:res}, we present the implications of the model for PNSs, and finally in \S~\ref{sec:discuss} we discuss the implications of our results for young neutron star populations.

\section{Model equations}
\label{sec:model}

There are now very sophisticated dynamo models that solve the PNS dynamo from the first principles \ie the magnetohydrodynamic equations \citep[see \eg][]{ray+20,mas+22}. These 2- and 3-dimensional simulations cannot be operated at realistic parameters (namely magnetic Reynolds numbers and magnetic Prandtl numbers) \citep{lan21} since this would require enormous computational power beyond the capability of the present resources \citep[but see][]{gui+22}. Thus, there is some room for reduced dynamo models, as we consider here.

As a simple description of the $\alpha-\Omega$ dynamo in PNS \citep{dun92,tho93}, we adopt the model proposed by \citet{wic+14} for the generation of magnetic fields in white dwarf mergers. This is a shear-driven dynamo model that effectively describes the turbulent dynamo process driven by differential rotation and convection, both of which undoubtedly operate in PNS. The new ingredients we add to the model are the required modifications owing to the shrinking radius of the PNS and the changing gravitational mass, and inclusion of viscous processes.

\subsection{The mean field dynamo}

In magnetohydrodynamics, the evolution of magnetic fields is governed by the magnetic induction equation
\begin{equation}
    \frac{\partial \mathbf{B}}{\partial t} = \nabla \times (\mathbf{U} \times \mathbf{B})+\lambda \nabla^2 \mathbf{B}
\end{equation}
where $\mathbf{U}$ is the velocity of the fluid and $\lambda$ is the magnetic diffusivity due to the electrical conductivity of the fluid.  This equation has been extensively employed for analyzing dynamos \citep{bran05,rin19, moff}.

In the mean-field dynamo model \citep{kra}, large-scale
magnetic fields are generated by small-scale turbulent flow. The application of the mean-field method to the induction equation leads to
\begin{equation}
\frac{\partial \mathbf{B}_0}{\partial t} = \mathbf{\nabla} \times \left(\mathbf{U}_{0} \times \mathbf{B}_{0} + \alpha \mathbf{B}_{0}-\beta \mathbf{\nabla} \times \mathbf{B}_{0} \right) + \lambda \nabla^{2} \mathbf{B}_{0}
\end{equation}
where subscript $0$ indicates the mean part of the fields and we will drop it in the following equations. Since the magnetic field can be written as the sum of the toroidal fields, $\mathbf{B} = \mathbf{B}_{\rm p} + \mathbf{B}_{\phi}$, the equation governing each component can be written as
\begin{align}
    \frac{\partial B_{\phi}}{\partial t} &= r\left(\mathbf{B}_{\rm p} \cdot \mathbf{\nabla} \Omega \right) +\lambda_{\rm e} \nabla^2 B_\phi~, \label{eq:tor}\\
    \frac{\partial B_{\rm p}}{\partial t} &= \alpha_{\rm t} \frac{\partial B_\phi}{\partial r} + \lambda_{\rm e} \nabla^2 B_{\rm p}~, \label{eq:pol}
\end{align}    
where we ignored several terms on the basis of being small. Here, $\alpha_{\rm t}$ is the turbulent transport coefficient. Its physical effect can be described as the twisting of the mean magnetic field lines by turbulence, resulting in the creation of magnetic field loops. These loops facilitate the conversion between the poloidal and toroidal magnetic fields. In the above equations,
$\lambda_{\rm e}$ denotes the effective magnetic diffusivity enhanced by turbulence. 

These equations motivate the reduced dynamo equations that we discuss below.

\subsection{Reduced dynamical equations for the fields}

According to \autoref{eq:tor}, the toroidal field, $B_{\phi}$, is generated from the coupling of the poloidal field with the differential rotation within the star (the first term on the right), known as the $\Omega$-effect, and is also reduced by diffusion due to turbulent magnetic diffusivity.
At a representative radius where the dynamo is the most effective, the evolution of the field can thus be written as
\begin{equation}
    \frac{d B_{\phi}}{dt} = q \Omega B_{\rm p} - \frac{B_{\phi}}{\tau_{\phi}}
\label{eq:toroidal}    
\end{equation}
where $q \equiv d\ln \Omega/d\ln r$ is the differential rotation \citep[see e.g.][]{bar+22} and $\tau_{\phi} = (\Delta R)^{2}/\lambda_{\rm e}$ is the diffusive time-scale, where $\Delta R$ is the size of the convective zone, but as we will discuss in \S~\ref{sec:angmom} its value will be set by field instabilities. The correspondence with \citet{wic+14} can be found by noting $q \sim \Delta \Omega / \Omega$. The effective magnetic diffusivity is defined by
\begin{equation}
	\lambda_{\rm e}= \frac{(\Delta R)^{2}}{\max \left(1, \frac{\Omega \tau_{\rm A}}{2 \pi}\right)    
		\tau_{\rm A}}
\end{equation}
(see \autoref{eq:difftime}). The convective shell thickness is set to $\Delta R = 10\,{\rm km}$ and kept constant throughout the evolution. With this choice, we ensure that upon collapse to a neutron star with radius $R=12\,{\rm km}$, the configuration is almost fully convective \citep{ray+22}.

According to \autoref{eq:pol}, the poloidal field is generated from the toroidal field by the $\alpha$-effect and is reduced by diffusion owing to the magnetic diffusivity. The effect is significant in the presence of cyclonic convection while it also relaxes on a time-scale $\tau_{\rm p}$ (see below) when the field configuration is unstable. Therefore, the evolution of the poloidal field is described by
\begin{equation}
\frac{d B_{\rm p}}{d t} = \alpha  \frac{B_{\phi}}{\tau_{\phi}} 
                    - \frac{B_{\rm p}}{\tau_{\rm p}} \qquad \mbox{for $R$ constant}
\label{eq:poloidalcntR}                    
\end{equation}
\citep{wic+14} where $\alpha=\tau_{\phi}\alpha_{\rm t}/\Delta R = \Delta R \alpha_{\rm t}/\lambda_{\rm e}$ is a factor that originates from the $\alpha$-effect that characterizes the efficiency of the dynamo action generating the poloidal field from the toroidal field. In order to determine the value of $\alpha$, we refer to $\alpha_{\rm t}$ from convective turbulence for slow and fast rotation limits \citep{pip+08,pip+12}. In our calculations, the influence of rotation is measured by Rossby number which is defined by $\text{Ro}=P/\tau_{\rm c}$ where $P$ is the rotational period of the star, $\tau_{\rm c}=h_{\rm p}/v_{\rm c}$ is the convective time-scale, the ratio of pressure scale-height over convective velocity \citep{tho93}. Using $\Delta R = \alpha_{\rm MLT}h_{\rm p}$ with $\alpha_{\rm MLT} \approx 1$ \citep{Kip+12} we found $\alpha_{\rm t}$ for slow rotation ($\text{Ro} \gg 1$) and fast rotation ($\text{Ro} \ll 1$) regimes as
\begin{equation}
	\alpha_{\rm t} = 
	\begin{cases}
		\Omega \Delta R & \text{ if } \text{Ro} \gg 1~, \\
		v_{\rm c} & \text{ if } \text{Ro} \ll 1~,
	\end{cases}
\end{equation}
where it is proportional to $\Omega$ for slow rotation, and independent of $\Omega$ for rapid rotation \citep{bran05}. Consequently, we define the dimensionless $\alpha$ by
\begin{equation}
	\alpha = \frac{\Delta R}{\lambda_{\rm e}} v_{\rm c} \min\left(1, \text{Ro}^{-1}\right)~.
	\label{eq:alpha}
\end{equation}
As it is expected, if Rossby number is small, then the dynamo action can be driven by a strong $\alpha$-effect \citep{bon+05}. 

The time-scale for the diffusion of the poloidal magnetic field is $\tau_{\rm p} = (\Delta R)^2/\lambda_{\rm e}$.

Considering that the radius of the PNS shrinks, the poloidal field would also increase by the conservation of magnetic flux ($\Phi$). If there was no dynamo action $\Phi = 4\pi R^2 B_{\rm p}$ with $d\Phi/dt=0$ would imply $\dot{B}_{\rm p} =-2B_{\rm p}\dot{R}/R$ where the dot ($\cdot$) denotes the time derivative. We thus replace \autoref{eq:poloidalcntR} with
\begin{equation}
    \frac{d B_{\rm p}}{d t} = \alpha \frac{B_{\phi}}{\tau_{\phi}} - \frac{B_{\rm p}}{\tau_{\rm p}} - 2 B_{\rm p}\frac{\dot{R}}{R} 
\label{eq:poloidal}
\end{equation}
where the last term stands for the field enhancement by flux conservation. The magnetic flux conservation, however, is most likely not satisfied exactly since the reconnection of the small-scale magnetic fields in the presence of the turbulence reduces the magnetic field \citep{mat+86,gold+92,eyi+13}. To account for the flux-loss due to magnetic reconnection, we introduced a term $-B_{\rm p}/\tau_{\rm rec}$ to the right hand side of \autoref{eq:poloidal} where $\tau_{\rm rec}$ is the reconnection time-scale which is defined by
\begin{equation}
	\tau_{\rm rec} = \frac{R}{\epsilon v_{\rm turb}}~.
\end{equation}
Here, $\epsilon$ is a parameter which determines the reconnection rate, the turbulent velocity $v_{\rm turb}$ is the convective velocity, $v_{\rm c}$. Thus, the poloidal field evolves with 
\begin{equation}
	\frac{d B_{\rm p}}{d t} = \alpha \frac{B_{\phi}}{\tau_{\phi}} - \frac{B_{\rm p}}{\tau_{\rm p,eff}} - 2 B_{\rm p}\frac{\dot{R}}{R}
\label{eq:ppoloidal}	
\end{equation}	
where 
\begin{equation}
	\frac{1}{\tau_{\rm p,eff}} = \frac{1}{\tau_{\rm p}} + \frac{1}{\tau_{\rm rec}}~. 
\end{equation}
We set $\epsilon=2.7 \times 10^{-4}$ so that $20\%$ of the magnetic flux is lost due to the reconnection, however, to investigate the impact of the flux-loss rate on the results, we also varied this parameter over a wide range.

\subsection{The stability of the fields}

It is well known that purely toroidal or poloidal fields are unstable and the unstable modes will dominate the loss of the fields by diffusion in unstable cases. 
Numerical simulations of \citet{bra09} suggest that for the toroidal and poloidal magnetic fields to be stable, they should satisfy the conditions
\begin{equation}
a \eta^2 < \eta_{\rm p} < 0.8 \eta
\label{eq:stability}
\end{equation}
where $a$ is the buoyancy factor, $\eta \equiv E/|U|$ is the ratio of magnetic energy to gravitational potential energy, and  $\eta_{\rm p}\equiv E_{\rm p}/|U|$ is the magnetic energy in the poloidal field scaled in the same way.
The buoyancy factor, $a$, limits the growth of the toroidal component in the dynamo process and is about $10$ for the main-sequence stars, and of order $10^{3}$ for neutron stars. \citep{bra09,wic+14}. Accordingly, we set the buoyancy factor as $a=10^{3}$ in most of the simulations, but also reveal its impact on the results by varying it over a wide range.

The $\Omega$-process for any differentially rotating or shearing system leads to an effective dynamo process coupled with the complementary $\alpha$-effect. The field, if it does not satisfy the above constraint given in \autoref{eq:stability}, may also decay with a timescale $\tau_{\phi}$ (see below).

According to \autoref{eq:stability} the toroidal field is subject to decay if $\eta_{\rm p}< a \eta^2$ \citep{bra09}. 
The total magnetic energy of the star is $E = B^2 R^3/6$ where we assumed the magnetic field to be uniform within the star. Similarly, the magnetic energy in the poloidal field is then $E_{\rm p} = B_{\rm p}^2 R^3/6$.
We scale these quantities with the gravitational potential energy of a uniform Newtonian star, $|U|= 3GM^2/5R$ to determine $\eta$ and $\eta_{\rm p}$ \citep{wic+14}.

The field instabilities, for a nonrotating star, operate on an Alfvén crossing-timescale $\tau_{\rm A}= R/v_{\rm A}$ where $R$ is stellar radius and $v_{\rm A} = B/\sqrt{4\pi \bar{\rho}}\simeq 10^5B_{12}~{\rm cm~s^{-1}}$ is the Alfvén velocity and $\bar{\rho}=3M/4\pi R^3\simeq 10^{14}~{\rm g~cm^{-3}}$ is the mean density of the star. In the presence of rotation, the growth rate of the instabilities is reduced by a factor of $\Omega \tau_{\rm A}/2\mathrm{\pi}$ \citep{pit85}. Thus, $\tau_{\phi}$ is given by
\begin{equation}
    \tau_{\phi}=
     \begin{cases}	
	\infty & \text { if } \eta_{\rm p}>a \eta^2~, \\
	\max \left(1, \frac{\Omega \tau_{\rm A}}{2 \mathrm{\pi}}\right) \tau_{\mathrm{A}} & \text { otherwise }~,
    \end{cases}\label{eq:difftime}
\end{equation}
where $\Omega$ is the stellar angular velocity. 

The time-scale for the decay of the poloidal field is given by
\begin{equation}
	\tau_{\rm p}=
	\begin{cases} 
		\infty & \text { if } 0.8 \eta>\eta_{\rm p}~, \\
		\max \left(1, \frac{\Omega \tau_{\rm A}}{2 \pi}\right)    
		\tau_{\rm A} & \text{ otherwise}~.
	\end{cases}
\end{equation}
\citet{wic+14} chooses $\alpha=1.52\times 10^{-4}$ for their simulations of white dwarf mergers.

\subsection{Angular momentum transport}\label{sec:angmom}

A PNS is born with some shear rate which is expected to get damped in magnitude over time and terminate the dynamo process eventually. We define this change as
\begin{equation}
    \frac{d q}{d t} = -\frac{q}{\tau}
\label{eq:differential}    
\end{equation}
where $\tau$ denotes its damping time-scale. One contributing factor to this change is the shrinkage of the PNS which amplifies the shear rate with a time-scale $\tau_{\rm I} = I/\left| \dot{I}\right|$ where $I = \gamma M R^2$ is the moment of inertia and we have taken $\gamma = 0.35$ for a PNS of $1.4 M_{\sun}$.

The shear rate, which drives the dynamo process, will be damped by the viscous processes \citep{sha00,mar+22} with a viscous timescale $\tau_\nu = R^{2}/\nu$ where $\nu$ is the internal viscosity. In scenarios involving a rapidly rotating PNS exhibiting a rotation period on the order of several milliseconds, the energy stored in rotation serves as a significant reservoir which can be effectively utilized when magnetic fields are present. In these conditions, turbulence due to magnetorotational instability (MRI) plays a crucial role in the angular momentum transport within stellar interiors \citep{aki+03} corresponding to an effective viscosity.
\begin{equation}
    \nu_{\rm MRI} = \frac{B_{\rm p} B_\phi}{4\pi\rho_{\rm crust} \left|q\right| \Omega}
\end{equation}
\citep{tho+05,ful+19,bar+22,mar+22} where $\rho_{\rm crust}=10^{13}\,{\rm g~{cm}^{-3}} \lesssim \bar{\rho}$ denotes the density of the PNS outer layers in which the MRI-driven turbulence is more relevant \citep{reb+22}.

The PNS, initially is opaque for neutrinos, and is expected to be convectively unstable due to entropy and/or compositional (leptonic) gradients \citep{eps79,bur88,bur92,mez+98,rob+12}. 
In fact, understanding the convective instability of the PNS gave rise to the original scenario invoked for the generation of magnetic fields of magnetars \citep{tho93}. 
Thus, convective dynamos has been studied extensively \citep{ray+20,ray+22,mas+20,mas+22,whi+22}.
\citet{mar+22} approximates the effective viscosity due to the convection as
\begin{equation}
\nu_{\rm conv} \simeq \tilde{\beta} v_{\rm c} \Delta R
\end{equation}
in which $\tilde{\beta}= 0.01$ represents the effectiveness parameter, $v_{\rm c}$ is the convective velocity and $\Delta R$ is the thickness of the convective shell. Relating the surface neutrino luminosity, $L_{\nu}$, to the convective flux, $F_{\rm conv}$, by $L_{\nu} = 10^{53}~{\rm erg/s} = F_{\rm conv} 4\pi r^2$ and referring $F_{\rm conv} \simeq \rho v_{\rm c}^3$,  the convective velocity can be written as
\begin{equation}
v_{\rm c} = 10^8~{\rm cm~s^{-1}} \left( \frac{L_{\nu}}{10^{53}~{\rm erg/s}}\right)^{1/3} \left(\frac{\rho}{\bar{\rho}} \right)^{-1/3}
\label{eq:convvel}
\end{equation}
\citep{mar+22}. Since the radius changes exponentially (see \autoref{eq:radius}) we have equated the neutrino luminosity to the time derivative of the gravitational potential energy
\begin{equation}
L_{\nu} = - \frac35 \frac{GM^2}{R^2} \dot{R}
\end{equation}
and allowed $L_{\nu}$ to change. A more sophisticated method can be found in \citet{suw+21}. Consequently, we have computed the internal viscosity through
\begin{equation}
    \nu = \nu_{\rm MRI} + \nu_{\rm conv}~,
\end{equation}
with the damping time-scale
\begin{equation}
	\frac{1}{\tau} = \frac{1}{\tau_\nu}-\frac{1}{\tau_I}~.
\end{equation}

\subsection{Evolution of the radius of the PNS} \label{sec:radius}

\begin{figure}
    \includegraphics[width=0.95\linewidth]{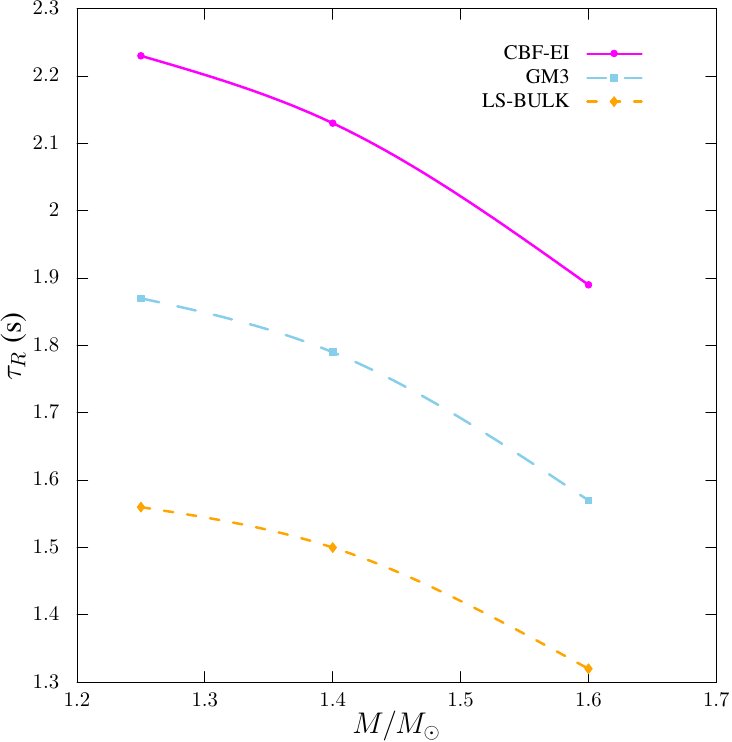}
    \caption{Dependence of the exponential decay time-scale of the radius of the PNS on mass and equation of state. We inferred this result from the data tables given in \citet{cam+17}. Different curves correspond to different EoS (see text).	\label{fig:tau_R}}
\end{figure}

\begin{figure*}
    \includegraphics[width=0.7\textwidth]{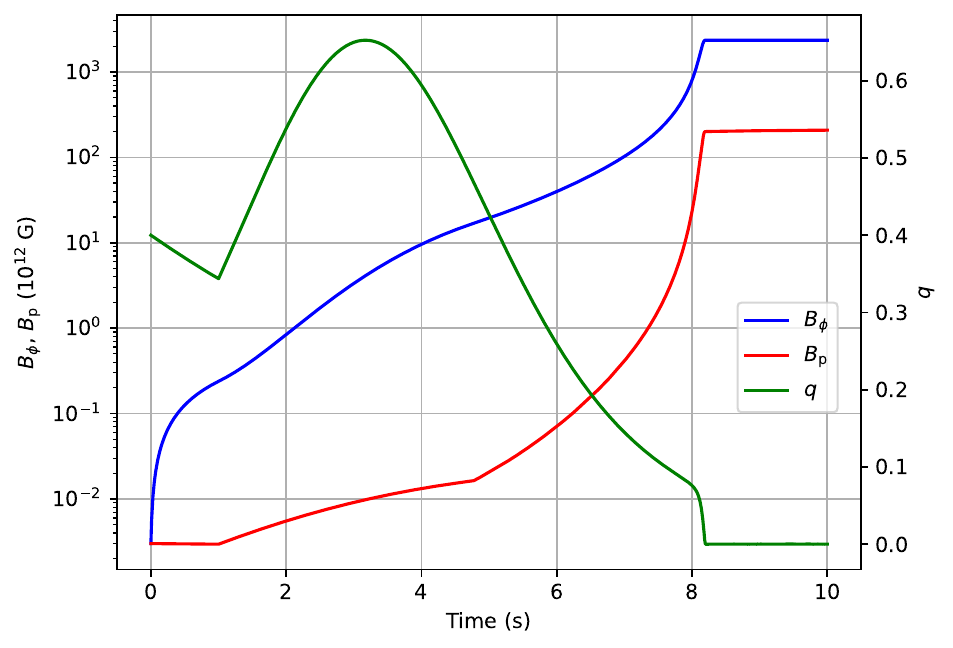}
    \caption{Evolution of the magnetic fields, $B_{\rm p}$ (red line), $B_{\phi}$ (blue line) and differential rotation, $q$ (green line). Here the initial conditions are $B_{\phi}(t=0) = B_{\rm p}(t=0) = 3\times 10^{9}\,{\rm G}$. $q_0 =0.4$ and $P_0 = 29.5\,{\rm ms}$ ($P_{\infty} = 2.5\,{\rm ms}$). We assumed $a=10^{3}$.}
\label{fig:fields}
\end{figure*}

During the PNS stage, the radius of the object shrinks from $R_0 \sim 40\,{\rm km}$ to $R_{\infty} \simeq 12\,{\rm km}$. The timescale for the collapse of the PNS depends on the mass of the PNS and the equation of state (EoS).
\citet{cam+17} modelled detailed evolution of the PNS for three different masses ($M=1.25,\,1.40,\,1.60\, M_{\sun}$) and three different EoS. 
The EoSs considered by \citet{cam+17} are (i) a nuclear many-body EoS obtained by correlated basis function theory with effective interaction (CBF-EI; \citet{ben17}), a mean field EoS by \citet{gle91} (GM3) and \citet{lat91} model (LS-bulk). We have seen that the evolution of the radius of the PNS, as obtained from the numerical results of \citet{cam+17}, can be fitted with an exponential decay
\begin{equation}
R(t) = (R_0 - R_{\infty}) \exp\left(-\frac{t-t_0}{\tau_R}\right) + R_{\infty}
\label{eq:radius}
\end{equation}
where $\tau_R=1.3-2.3$~s depending on the EoS and the mass of the PNS as shown in \autoref{fig:tau_R}. The evolution starts after $t_0 \sim 1$~s and we assumed, for simplicity that, for $0<t<t_0$, the radius is constant, $R=R_0$, which is obviously not correct but has little effect on our results since this is a short time interval. These are consistent with the results of \citet{nag+20}.
We note that the above equation is the solution of the differential equation $\dot{R} = -(R- R_{\infty})/\tau_R$
and we have used this form whenever it simplifies calculations.

\subsection{Evolution of the gravitational mass of the PNS}

Since PNS are relativistic objects there is a difference between the gravitational mass and the baryonic mass depending on the compactness, $GM/Rc^2$. Although the number of baryons and hence the baryonic mass remains constant, the gravitational mass decreases as the radius shrinks \citep{cam+17}. We have seen that the evolution of the gravitational mass can also be fitted with an exponential function
\begin{equation}
M(t) = (M_0 - M_{\infty}) \exp\left(-\frac{t-t_0}{\tau_M}\right) + M_{\infty}
\label{eq:mass}
\end{equation}
with $3.4~{\rm s}<\tau_M<4.5~{\rm s}$. We thus considered the mass of the PNS evolves according to $\dot{M}=-(M-M_{\infty})/\tau_M$, though it only leads to about $10\%$ difference in mass. We again assumed that the mass is constant for $0<t<t_0$ for the same reason as referred to in \S~\ref{sec:radius}.

\subsection{Evolution of the spin angular velocity}

The angular velocity of the shrinking PNS needs to increase due to the conservation of angular momentum as
\begin{equation}
\frac{d\Omega}{dt} = - \frac{\dot{I}}{I} \Omega~.
\label{eq:omega}
\end{equation}
Here, the moment of inertia will change both due to change in the radius and gravitational mass.
The external torques on the object due to electromagnetic emission, gravitational waves and neutrinos will try to spin it down. It is likely that the neutrino-driven PNS winds \citep{bur+95,jan96} lead to even stronger external spin-down torques \citep{buc+06,met+07}. We first added these effects, but we have also seen that the change in the spin frequency due to the change in the moment of inertia dominates the change due to these external torques. We thus, for simplicity, do not include the external torques during the short time interval we consider in this work.  

\subsection{Termination of the dynamo action \label{sec:terminate}}

Several processes contribute to the termination of the dynamo process, the main one being the loss of leptonic gradients reducing the convective processes.

Initially, the PNS is opaque and the neutrino cooling proceeds by diffusion until the temperature drops to $\sim 10^{10}\,{\rm K}$ (0.86 MeV). This happens at around 40 s. Once the stellar matter becomes neutrino-transparent ($T \simeq 10^{10}\,{\rm K}$), any gradients sustaining the convection will flatten and this will quench the $\alpha$-process. The dynamo process, thus, proceeds for about $40\,{\rm s}$ \citep{ray+20} which is sufficient to achieve saturation of the poloidal component with regard to the $\alpha$-effect. 

Rapid cooling of the neutron star as it becomes neutrino-transparent will lead to the solidification of a crust \citep{bur86, glen2000}. An effective viscosity approaching infinity damps the differential rotation to that of a rigid body, thus effectively suppressing the sources of turbulence \citep{tal+97, mey2000}.

\section{Results}
\label{sec:res}

We have numerically solved Equations \eqref{eq:toroidal}-\eqref{eq:differential} with the Runge-Kutta method.
We set the initial parameters as $M_0=1.55 M_{\sun}$ and $R_0=40\,{\rm km}$ as suitable for a PNS and $M_{\infty} = 1.46 M_{\sun}$ and $R_{\infty}=12$~km which are appropriate for a cold neutron star. The initial angular velocity is derived from the conservation of angular momentum such that the PNS will have $P_{\infty}\simeq 3~{\rm ms}$ when it settles. We set the seed fields to $B_{\rm p}(t=0) = 3\times 10^{9}\,{\rm G}$ and $B_{\phi}(t=0) = 3\times 10^{9}\,{\rm G}$, representing the flux-freezing estimate under full flux conservation as discussed in \S~\ref{sec:intro}. Departures from ideal flux-freezing are then included through a reconnection-driven flux loss. We have also studied the implications of changing these parameters.

In \autoref{fig:fields} we show the evolution of the poloidal and toroidal magnetic fields and the shear rate of the star for a PNS initially rotating with $P_0 = 29.5\,{\rm ms}$. The PNS transforms into a magnetar approximately $10\,{\rm ms}$ after the dynamo process, with a rotational period of $P_\infty = 2.5\,{\rm ms}$. We note that the dynamo process can produce various types of NSs, like normal pulsars or low-B magnetars, for PNSs that initially have longer rotational periods as our following analysis, showing the dependence on initial parameters, implies. The initial evolution of the fields and differential rotation matches with that given in \citet{wic+14} if the radius is kept constant and viscosity is neglected. In this regime, the toroidal field grows in time from the seed field until attaining its saturation value, $B_{\phi,\infty}$. Simultaneously, the poloidal field grows monotonically and approaches the value at which it will saturate, $B_{\rm p,\infty}$, due to the non-linear feedback of $B$ on the shear.

We note that the poloidal field is one order of magnitude smaller than the toroidal field. The poloidal field will be enhanced upon the collapse of the PNS to the typical size of a neutron star (we assumed $80\%$ of the flux is conserved). The toroidal field also is affected by the collapse, though not by the flux conservation, but indirectly, according to the scaling relation of $B \propto \rho^{2/3}$ and through the stability criteria given in \autoref{eq:stability}. The gravitational potential energy $|3GM^2/5R|$ will also be enhanced by a factor of $40\,{\rm km}/12\,{\rm km} \simeq 3.3$. 

The shear rate attains a maximum value during the contraction phase of the PNS. Subsequently, it undergoes a gradual decline to zero, consistent with the theoretical expectations. This arises from the significant increase in the internal viscosity which suppresses the differential rotation, and consequently reduces the shear rate.

\subsection{Dependences on the parameters\label{sec:dependences}}

We have investigated the effect of the parameters of the model on the resulting fields.

In the preceding subsection, we have chosen specific parameters for the PNS. Specifically, we set the initial mass and radius of the PNS as $M_0=1.55 M_{\sun}$ and $R_0=40\,{\rm km}$, respectively. We assumed the initial shear rate is $q_{0} = 0.4$. Additionally, we defined $a = 10^{3}$. We adopt  $B_{\rm p}(t=0) = 3\times 10^{9}\,{\rm G}$ and $B_{\phi}(t=0) = 3\times 10^{9}\,{\rm G}$, i.e. the value obtained for perfect flux conservation. Possible flux loss is captured by the reconnection term in the evolutionary equation of the poloidal field, with $\tau_{\rm rec}=R/\epsilon v_{\rm turb}$ controlling the fractional flux loss. We set $20\%$ of the flux is lost by reconnection process.

Here, we study the implications of changing all these parameters.

\subsubsection{Dependence on the initial value of the shear rate, $q_{0}$}

\begin{figure}
    \includegraphics[width=0.95\linewidth]{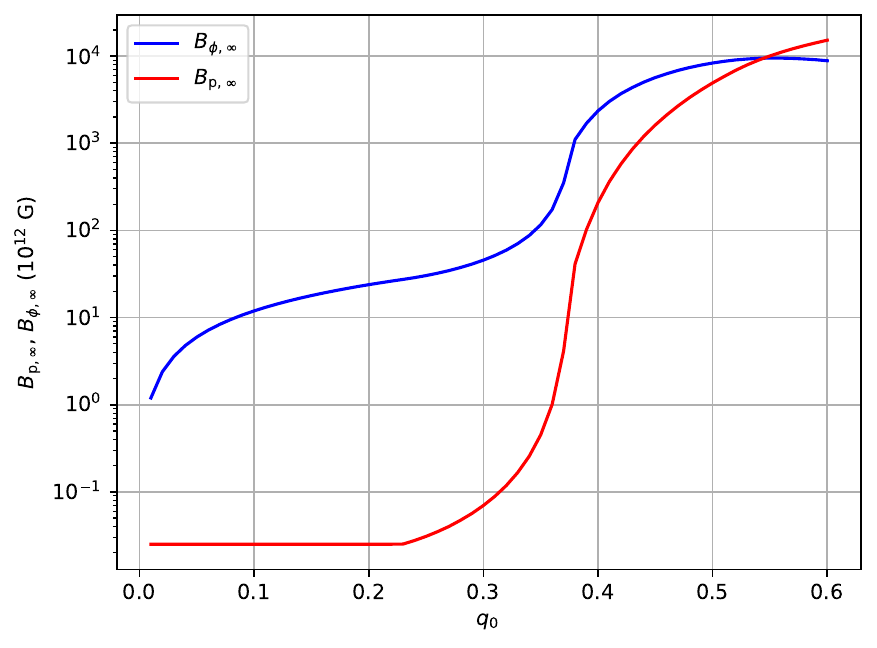}
    \caption{Effect of the initial shear rate, $q_{0}$, on the saturation value of magnetic fields, $B_{\rm p, \infty}$ (red line) and $B_{\phi, \infty}$ (blue line). Here the initial fields are $B_{\phi}(t=0) = 3\times 10^{9}\,{\rm G}$ and $B_{\rm p}(t=0) = 3\times 10^{9}\,{\rm G}$. We assumed $a=10^{3}$.}
\label{fig:q0}
\end{figure}

We analyzed the effect of the initial shear rate, $q_{0},$ of the PNS on the final values of the magnetic fields. We varied it within the range of $0.01-0.60$. Our analysis demonstrates that the saturation value of the toroidal field $B_{\phi, \infty}$ depends strongly on the initial shear rate, $q_{0}$. However, it exerts a negligible influence on $B_{\rm p, \infty}$ up to a specific threshold, $q_{0}=0.23$, as illustrated in \autoref{fig:q0}. Higher initial shear rate yields significantly larger $B_{\phi, \infty}$ since larger $q_{0}$ makes the source of the $\Omega$-effect stronger at early times. In addition, since the MRI viscosity scales inversely with the shear rate, larger $q_{0}$ reduces $\nu_{\rm MRI}$, increases the viscous time-scale and therefore slows the decay of $q$. As a result, source of the $\Omega$-effect remains active for longer which leads to a larger production of the toroidal field and higher $B_{\phi, \infty}$. The enhanced toroidal field feeds back into the poloidal production modestly raising $B_{\rm p, \infty}$ since the generation of the poloidal field depends on $q_0$ only indirectly via the $\Omega$-effect. In the regime where $q_0 < 0.23$, the saturation value of the poloidal field is $B_{\rm p, \infty}=2.5 \times 10^{10}\,{\rm G}$ which is the field strength that would be inherited by $80\%$ flux conservation without any dynamo operation. Since the poloidal field is generated from the decaying toroidal field, the $\alpha$-effect must be triggered by the magnetic instabilities (see \autoref{eq:ppoloidal}). Thus, we can conclude that in the regime with  $q_0 < 0.23$, where the poloidal field remains insensitive to $q_0$, it is not enough to trigger the $\alpha$-effect, and the only mechanism that amplifies the poloidal field is the flux conservation. Results also show that the saturation values of both fields are equal such that $B_{\phi, \infty} = B_{\rm p, \infty} = 9.5 \times 10^{15}\,{\rm G}$ at $q_0 = 0.54$. Once the initial shear rate exceeds this threshold, poloidal production dominates over toroidal production and the system switches to $B_{\rm p, \infty} > B_{\phi, \infty}$ region. Although dynamo arguments might suggest that higher shear rate should lead to a toroidal dominated configuration, the combination of shear dependent dissipation and magnetic instability thresholds in our model limits the growth of $B_\phi$ while still allowing an efficient operation of $\alpha$-effect. In this sense, the higher $q_0$ regime corresponds to a mixed field configuration in which magnetic instabilities limit the growth of the toroidal field, while an efficient $\alpha$-effect and further amplification of $B_{\rm p}$ by flux compression make the configuration poloidal-dominated, in agreement with stability studies indicating upper limits on the toroidal energy fraction in the mixed equilibria \citep{bra09, cio+10, lan+13}.

\subsubsection{Dependence on the flux loss rate}

\begin{figure*}
    \includegraphics[width=0.49\textwidth]{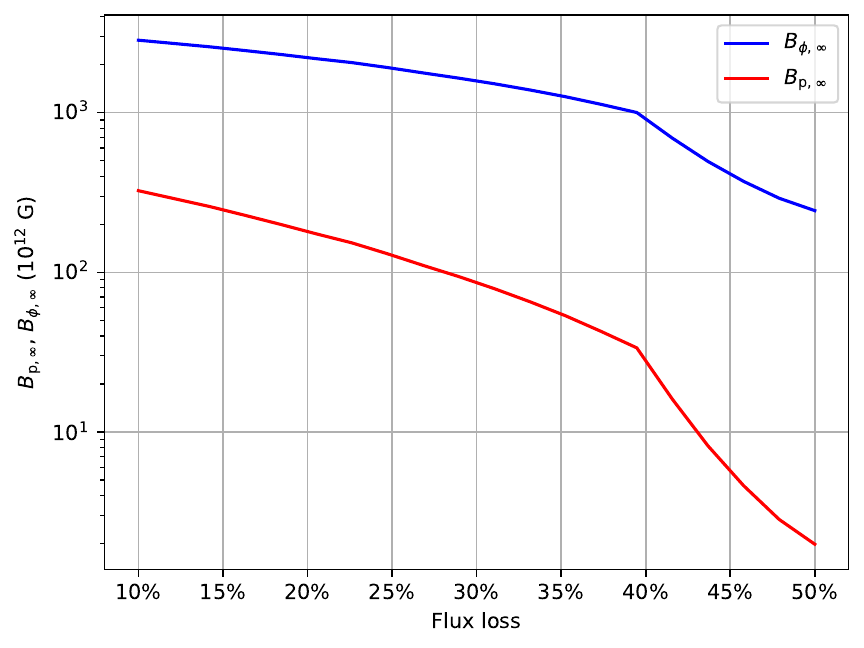}
    \includegraphics[width=0.49\textwidth]{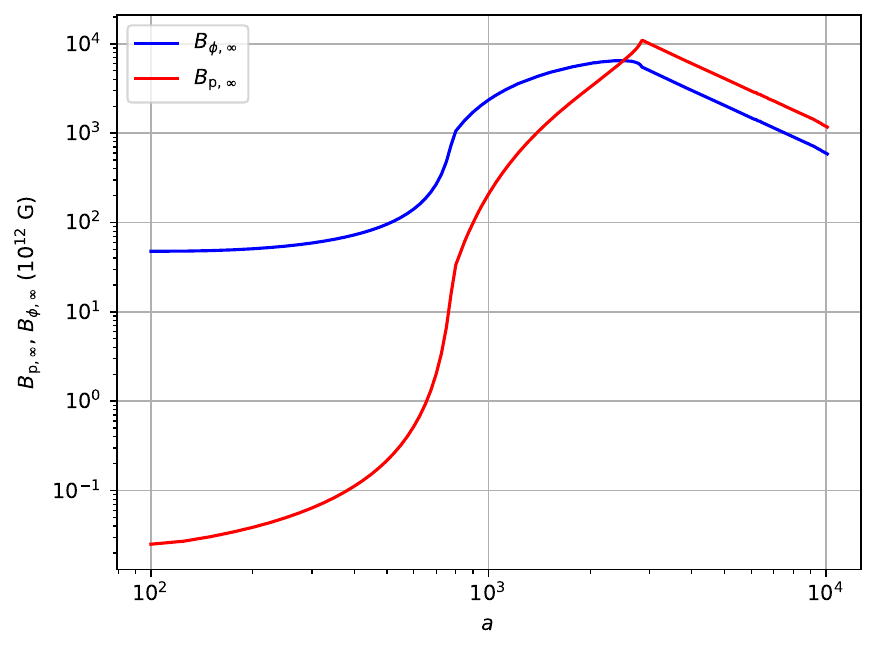}
    \caption{Here $B_{\phi}(t=0) = 3\times 10^{9}\,{\rm G}$, $B_{\rm p}(t=0) = 3\times 10^{9}\,{\rm G}$ and $q_0 = 0.4$. (Left panel) Effect of the flux loss rate on the saturated value of the poloidal field, $B_{\rm p, \infty}$  (red line) and the toroidal field, $B_{\phi, \infty}$ (blue line) with $a=10^3$. (Right panel) Effect of buoyancy factor, $a$, on the saturated value of the poloidal field, $B_{\rm p, \infty}$  (red line) and the toroidal field, $B_{\phi, \infty}$ (blue line).}
\label{fig:eps-a}
\end{figure*}

We also analyzed the results of simulations in which the flux loss rate is varied in the interval $10\%-50\%$, implemented by appropriate choices of $\epsilon$. As the flux loss fraction increases, both saturated fields decline monotonically, once the flux loss fraction exceeds about $40\%$, they begin to decrease much more rapidly with further flux loss as shown in the left panel of \autoref{fig:eps-a}. The poloidal component is more sensitive since the reconnection directly increases its effective loss rate whereas the toroidal component responds only through the weakened source term.

\subsubsection{Dependence on the buoyancy parameter, $a$}

We show the effect of the buoyancy factor, $a$, on  $B_{\rm p,\infty}$ and $B_{\phi,\infty}$ by varying it within the interval of $10^{2}-10^{4}$ in the right panel of \autoref{fig:eps-a}. We found that for higher buoyancy parameter up to the certain values of $a \simeq 2.5 \times 10^3$ and $a \simeq 2.9 \times 10^3$, $B_{\phi,\infty}$ and $B_{\rm p, \infty}$ are stronger respectively. In this regime, $B_{\phi,\infty}$ remains larger than the poloidal one. This follows from the $a$ dependence of the toroidal loss. When $\eta_{\rm p}>a \eta^2$, the toroidal damping is switched off; and this condition is more readily satisfied for smaller $a$. When toroidal loss is closed, poloidal field generation by $\alpha$-effect is ineffective; however, flux compression can still feeds it. Hence, the poloidal amplitude stays smaller than the toroidal one. Beyond this critical values, further increase in $a$ leads to smaller saturated fields since the loss channel opens more easily and damping strengthens.

\subsubsection{Dependence on the initial mass of the PNS, $M_{0}$}

\begin{figure*}
	\includegraphics[width=0.49\textwidth]{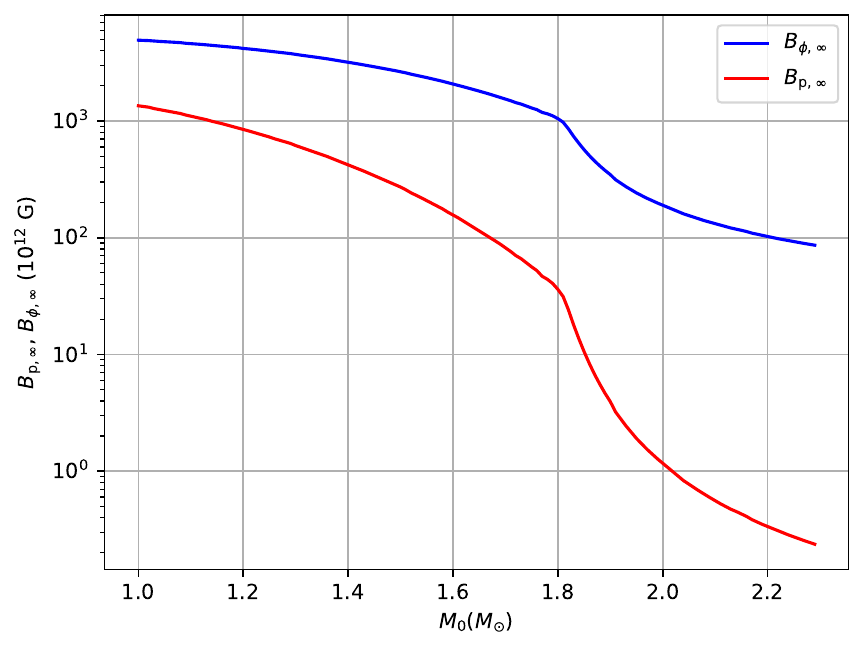}
    \includegraphics[width=0.49\textwidth]{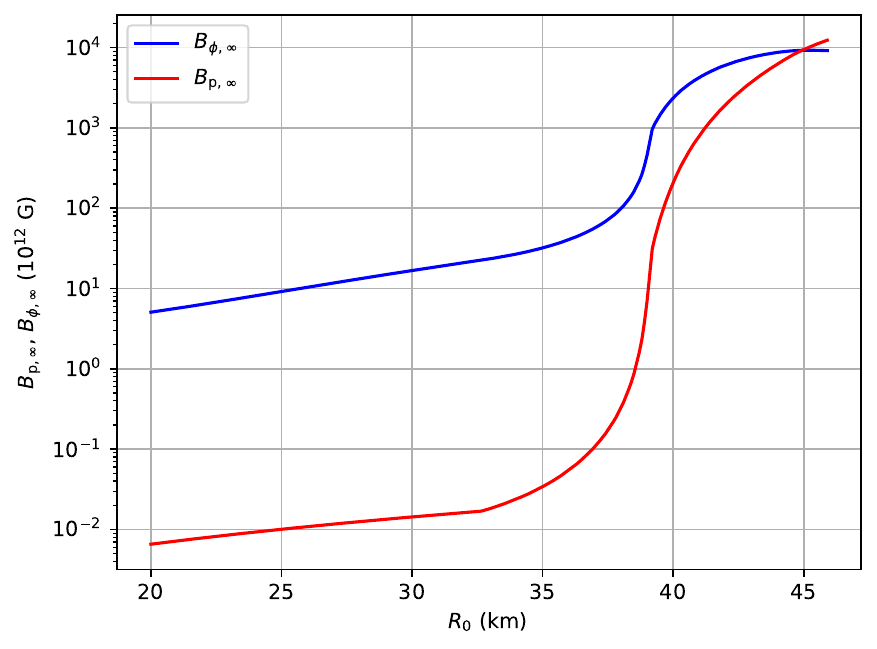}
	\caption{Here $B_{\phi}(t=0) = 3\times 10^{9}\,{\rm G}$, $B_{\rm p}(t=0) = 3\times 10^{9}\,{\rm G}$, $q_0 = 0.4$, $a = 10^{3}$. (Left panel) Effect of the initial mass of the PNS, $M_{0}$, on the saturated value of the poloidal field, $B_{\rm p, \infty}$ (red line) and the toroidal field, $B_{\phi, \infty}$ (blue line) with the initial radius of $R_{0} = 40\,{\rm km}$. (Right panel) Effect of the initial radius of the PNS, $R_{0}$, on the saturated value of the poloidal field, $B_{\rm p}$ (red line) and the toroidal field, $B_{\phi}$ (blue line) with the initial mass of $M_{0} = 1.55 M_{\sun}$.}
\label{fig:M-R}
\end{figure*}

We have investigated the effect of the initial mass of the PNS, $M_{0}$, on $B_{\rm p,\infty}$ and $B_{\phi,\infty}$ (see left panel of \autoref{fig:M-R}). We choose an initial mass range of $1-2.2 M_{\sun}$ which is appropriate for a PNS. Our findings indicate that both fields exhibit a decreasing trend with the initial mass of the PNS. For initial masses $M_0 > 1.8 M_{\sun}$, the decrease in the saturation values with initial mass is observed to become more rapid. The initial angular velocity is set by the conservation of angular momentum $(\Omega_0 \propto M_0^{-1})$, then higher-mass progenitor PNSs experience less efficient $\Omega$-effect at early times. Subsequently, $\alpha$-effect is also initially weaker for those PNSs. This fact can be understood as higher Rossby number regime $(\text{Ro} \propto \Omega^{-1})$ exhibit weaker $\alpha$-effect as it is expected \citep{tho93, bon+05}.  Nevertheless, it is noteworthy that the enhancement in the saturated toroidal field strength is comparatively less pronounced than that observed in the saturated poloidal field strength.

\subsubsection{Dependence on the initial radius of the PNS, $R_{0}$}

We further explored how the initial radius of the PNS, $R_{0}$, impacts $B_{\rm p, \infty}$ and $B_{\phi,\infty}$. Our analysis was conducted by adjusting $R_{0}$ in the range of $20-45\,{\rm km}$ which is the usually adopted range of radii for PNSs. Our findings indicate that when the PNS undergoes a dynamo mechanism with a larger initial radius, the magnetic fields reach higher intensities at the point of saturation. This trend follows directly from the structure of the evolution equations. The final radius of the star is fixed at $R_\infty=12\,{\rm km}$ in this examination, which provides a higher shrinkage rate if the initial radius of the PNS is larger. Consequently, the early growth of the poloidal field  due to flux conservation is stronger. Moreover, the viscous time-scale initially increases with larger $R_0$, since the radius depencence in $\tau_\nu$ dominates over the modest increase in the convective viscosity. Accordingly, the shear is damped less rapidly, and the source of the $\Omega$-effect stays active for longer. When the initial radius is $R_0\sim45\,{\rm km}$, the field strengths at saturation become equal at $B_{\phi,\infty} = B_{\rm p,\infty} = 9 \times 10^{15}\,{\rm G}$, and for larger radii the saturation value of the poloidal field becomes greater than that of the toroidal field. In this regime, both $\Omega$- and $\alpha$-sources initially weaken, but field amplification by flux conservation is strengthened leading to a poloidal-dominated configuration. The right panel of \autoref{fig:M-R} illustrates that the initial radius of the PNS significantly influences its ultimate magnetic fields.

\subsubsection{Dependence on the initial value of the fields, $B_{\rm p,0}$ and $B_{\phi, 0}$}

\begin{figure*}
    \includegraphics[width=0.49\textwidth]{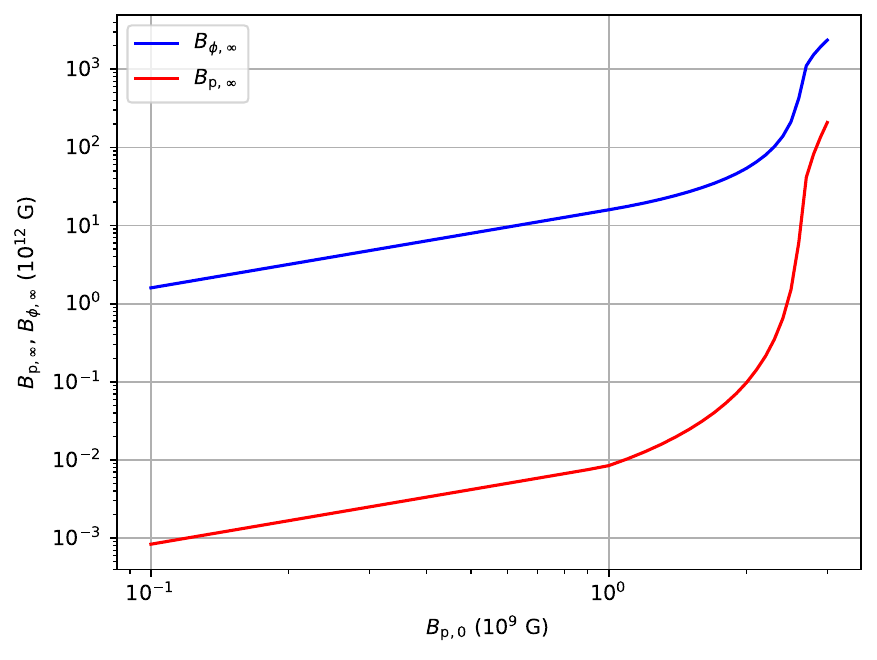}
    \includegraphics[width=0.49\textwidth]{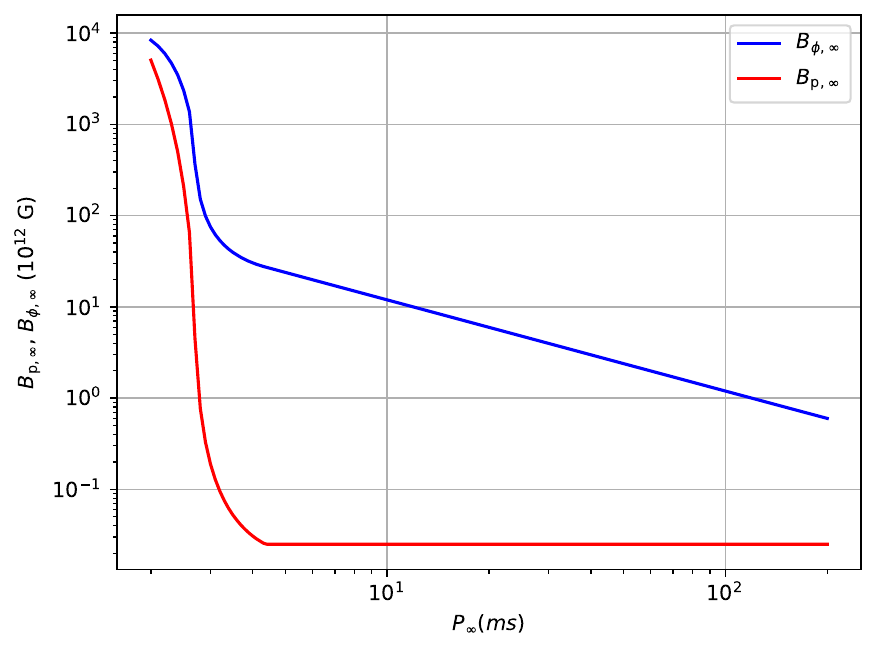}
    \caption{Effect of the initial poloidal field $B_{\rm p}(t=0)$ (left panel) and the final rotational period $P_{\infty}$ (right panel) on the saturation values of the poloidal field, $B_{\rm p, \infty}$ (red line), and toroidal field $B_{\phi, \infty}$ (blue line). Here $a=10^{3}$ and $q_0=0.4$. For the left panel, the initial value of the toroidal field is $B_{\phi}(t=0) = 3\times 10^{9}\,{\rm G}$. The nitial fields are $B_{\phi}(t=0) = 3\times 10^{9}\,{\rm G}$, $B_{\rm p}(t=0) = 3\times 10^{9}\,{\rm G}$ for the right panel.}
\label{fig:Bp0-Pinf}
\end{figure*}

We also checked the dependence of $B_{\rm p, \infty}$ and $B_{\phi,\infty}$ on the initial values of the poloidal and toroidal fields, $B_{\rm p,0}$ and $B_{\phi, 0}$, respectively. We investigate the results for both seed fields within the interval of $1 \times 10^{8}\,{\rm G}$ to $3 \times 10^{9}\,{\rm G}$. The results are shown in the right panel of \autoref{fig:Bp0-Pinf}. It is observed that the saturated fields show a strong dependence on $B_{\rm p,0}$ while they have nearly no dependence on $B_{\phi, 0}$. 

Results indicate that the fields at saturation increase with the seed poloidal field $B_{\rm p,0}$. As $B_{\rm p,0}$ increases, the $\Omega$-effect, whose source is always active, initially strengthens since its source is proportional to the shear acting on the existing poloidal field, resulting in stronger toroidal field generation. Recall that the $\alpha$-effect generates poloidal field from the decaying toroidal field in this model. Thus, more efficient toroidal field generation directly feeds the poloidal field generation as well. With stronger $B_{\rm p,0}$, the amplification of the poloidal field by flux compression also increases. Therefore, we observe stronger fields at saturation as $B_{\rm p,0}$ increases. On the other hand, we see that the initial value of the toroidal field is not a determinative factor at which levels the magnetic field components will be saturated in the dynamo process. $B_{\phi, 0}$ is initially used by the $\alpha$-effect whose source must be triggered by the instabilities in contrast to the source of the $\Omega$-effect. Thus, our findings show that even a larger seed toroidal field does not effect the faith of the field components since the $\alpha$-effect is not initially triggered. In short, the initial value of the toroidal field is not important in setting the final values of the fields in the dynamo process, while stronger fields can be generated if the poloidal field inherited by flux conservation is high.

\subsubsection{Dependence on  the final rotational period, $P_{\infty}$ \label{sec:P_infty}}

We analyzed the dependence on the final rotational period of the PNS following the settling to the conventional neutron star radius, $R=12~{\rm km}$. We perform the analysis for $P_{\infty}$ values that fall within the range of $2-200~{\rm ms}$. The results depicted in the left panel of \autoref{fig:Bp0-Pinf} demonstrate that, for the shortest spin periods like $P \sim 2-3~{\rm ms}$, a PNS transitions into a magnetar with magnetic fields up to $B_{\rm p} \sim 10^{15}\,{\rm G}$ and $B_{\phi} \sim 10^{16}\,{\rm G}$. It is also observed that it may ultimately become a high-magnetic field pulsar possessing a strong toroidal magnetic field $\sim10^{14}\,{\rm G}$. Our results also show that at the end of the $\alpha-\Omega$ dynamo, the PNS can become a normal pulsar with $B_{\rm p} \sim 10^{12}\,{\rm G}$ and $B_{\phi} \sim 10^{14}\,{\rm G}$. This finding confirms the conventional wisdom that the shorter the rotational period of the neutron star, the stronger its toroidal field strength. 
Accordingly, those PNSs with slightly longer rotational periods, will become low-B magnetars \citep[see \eg][]{rea+13,rea+14} and have toroidal fields exceeding those of normal pulsars by an order of magnitude, reaching $\sim 10^{15}\,{\rm G}$. This was recently confirmed by \citet{igo+25} whose numerical simulations demonstrate that the burst activities of low-B magnetars can be explained through a magnetic field structure generated by a dynamo process.

Our results suggest that the final value of the poloidal field becomes independent of $P_{\infty}$ beyond $4.4\,{\rm ms}$, where the poloidal field is $2.5 \times 10^{10}\,{\rm G}$, i.e.\ that inherited by $80\%$ flux conservation. Interestingly, our results indicate that the toroidal field linearly decreases with the initial period in this regime. This regime produces normal pulsars with relatively small magnetic fields, $B_{\rm p} \sim 10^{10}-10^{11}~{\rm G}$. Numerical simulations do not study this regime where the dynamo process sets the toroidal field.

Central compact objects (CCOs) have spin periods in the range 100-400 ms and their relatively small magnetic fields $B_{\rm p} \sim 10^{10}~{\rm G}$ imply that these are very close to their initial spin periods \citep{got07,hal+07,got09,del17}. Our results suggest that, although their poloidal fields are no different than what they inherited from their progenitor by flux conservation, their toroidal fields are shaped by the dynamo process.

\begin{figure}
    \includegraphics[width=0.95\linewidth]{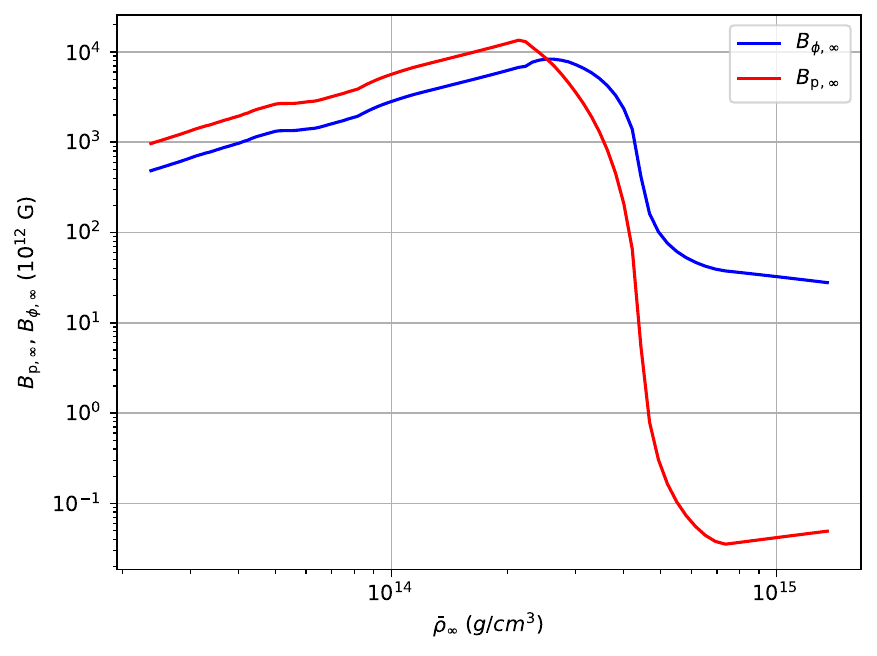}
    \caption{Effect of the final mean density of the star, $\bar{\rho}_{\infty}$, on the saturated poloidal field, $B_{\rm p, \infty}$ (red line) and toroidal field, $B_{\phi, \infty}$ (blue line). Here the initial fields are $B_{\phi}(t=0) = 3\times 10^{9}\,{\rm G}$, $B_{\rm p}(t=0) = 3\times 10^{9}\,{\rm G}$ and the initial shear rate is $q_{0}=0.4$. We assumed $a=10^{3}$.}
\label{fig:rho_inf}
\end{figure}

\subsubsection{Dependence on the final mean density, $\bar{\rho}_{\infty}$}

We also examined the influence of the final average density of the star following the dynamo process on the saturation levels of the fields. In \autoref{fig:rho_inf}, we present the findings, indicating that the saturation values of the fields increase as the average density grows, up to a certain value. In this regime, the combination of favourable instability criteria and shear allows the dynamo loop to operate more efficiently. Differential rotation and convection are still strong enough so that the system has enough time to build up comparatively large saturated fields. However, after exceeding this threshold, the fields begin to show an inverse relationship and decrease as the average density increases. As the compactness increases, the differential rotation becomes strongly dissipated, the operation of the magnetic instabilities become more difficult with longer $\tau_{\rm A}$, and the $\alpha$-effect weakens due to the reduced convective velocity and less favourable instability thresholds. As a result, dynamo operates less efficiently leading to smaller saturation fields at the highest densities. In addition, this reduction also accelerates beyond an average density value.

\section{Discussion}
\label{sec:discuss}
We studied the $\alpha - \Omega$ dynamo process, which is largely assumed to operate during the PNS phase. This dynamo mechanism is recognized for its ability to produce strong magnetic fields through the combined effects of vigorous differential rotation and convective motions; hence, it is regarded as a potential source for the magnetic fields of the magnetars.

We used a simple one-dimensional model originally introduced by \citet{wic+14} for mergers of white dwarfs. We additionally included the effects of fundamental evolutionary processes like the shrinkage of the PNS and the resulting reduction in the gravitational mass. We also consider magnetorotational and convective instabilities which play an important role in the dynamo process contributing to the internal viscosity. The simplicity of the model allows us to study the full episode of the dynamo and make a parameter scan to see the dependencies.

We emphasise that the background PNS evolution used in this paper is based on one-dimensional models without mixing-length convection. We use the results of \citet{cam+17} only to parametrise the global evolution of the PNS radius and gravitational mass through \autoref{eq:radius} and \autoref{eq:mass}. These one-dimensional cooling models do not include convective energy and lepton transport via mixing-length theory. More recent calculations that solve the quasi-static PNS evolution including convection within MLT \citep[\eg][]{pas+22} show that convective motions can modify the contraction history. Our dynamo equations are, however, formulated in terms of global quantities, and the convective velocity inferred from energetics (\autoref{eq:convvel}). Therefore, we expect the main trends with the initial rotation, shear and buoyancy parameters to remain qualitatively unchanged. A more consistent treatment, in which the convective fluxes and turbulent transport coefficients are taken directly from multi-group neutrino-radiation-hydrodynamics simulations, would be an important next step.

Following a recent argument by \citet{lan21} that the magnetic fields of neutron stars obtained through flux conservation are small, we considered the possibility that normal pulsars also pass through the same dynamo process, albeit with larger initial periods.
We considered a specifically wide range of initial spin periods reaching up to 200 ms.

Our findings indicate that, under different initial conditions of convective flow and rotation rate, PNSs can not only have magnetic fields evolving into magnetar level strengths ($\gtrsim 10^{14}\,{\rm G}$), but also into the field strength of distinct sub-populations that may be observed as low-B magnetars, high-field pulsars and normal pulsars. This diversity arises from the sensitivity of the dynamo process to the initial conditions and to the internal structural parameters of the star. Our study confirms that PNSs with very rapid rotation rates are particularly prone to magnetar formation, whereas those with more moderate spin rates can evolve into high-magnetic field pulsars or even normal pulsars. In this picture, only central compact objects would not pass through a dynamo process, since their fields can easily be explained with flux conservation. This is also consistent with the initial periods estimated for these objects: $P_0 \simeq 100\,{\rm ms}$. Such slow periods are unlikely to trigger dynamo action within the neutron star. We found that, although the poloidal fields of CCOs are set by flux conservation, their toroidal fields are shaped by the $\Omega$-effect (\S~\ref{sec:P_infty}).

All neutron stars may have some small-scale fields. However, it appears that the relative strength of the small-scale fields of CCOs with their dipole fields is greatest among young pulsars, a result of their relatively small dipole fields. In general, the dynamo action organizes the fields by converting the small-scale fields into dipole and toroidal fields.  As such, the relatively strong small-scale fields of CCOs \citep{sha12,gou+20,alf+22} could be attributed to the fact that they have not gone through a genuine dynamo process.

Our results can be qualitatively related to existing three-dimensional simulations of convective and MRI-driven dynamos in PNSs. Global simulations by \citet{ray+20,ray+22} and \citet{reb+21,reb+22} typically find that rapidly rotating PNSs are able to generate strong large-scale magnetic fields with a significant toroidal component, whereas more slowly rotating models produce weaker fields. This trend with rotation is the same as in our one-zone model where very short initial periods favour magnetar-strength configurations, while more moderate spins lead to fields in the high-B and ordinary pulsar range. A detailed, quantitative comparison of our saturated field strengths and effective transport coefficients with these three-dimensional simulations is left for future studies.

Consequently, our examination of the $\alpha-\Omega$ dynamo process suggests that this model, initially proposed for magnetars, has a reasonable parameter space that can address young neutron star families.

\section*{Acknowledgements}
We thank M.~Ali Alpar, Emre I{\c s}{\i}k, Erbil G{\"u}gercino{\u g}lu and Shotaro Yamasaki for their valuable comments on an early version of the manuscript.

\section*{Data availability}
No new data were analysed in support of this paper.

\footnotesize{
\bibliographystyle{mnras}
\bibliography{dynamo.bib} 
}

\end{document}